\documentclass{webofc}
\usepackage[varg]{txfonts}   
\begin{document}
\title{$\beta$-decay spectroscopy of neutron-deficient nuclei}
%

\author{\firstname{Sonja E. A.} \lastname{Orrigo}\inst{1}\fnsep\thanks{\email{Sonja.Orrigo@ific.uv.es}} \and
        \firstname{Berta} \lastname{Rubio}\inst{1} \and
        \firstname{William} \lastname{Gelletly}\inst{2}
}

\institute{Instituto de F{\'i}sica Corpuscular, CSIC-Universidad de Valencia, E-46071 Valencia, Spain 
\and
           Department of Physics, University of Surrey, Guildford GU2 7XH, Surrey, UK 
          }

\abstract{%
A systematic study of the $\beta$-decay of neutron-deficient nuclei has been carried out and has provided spectroscopic information of importance for both nuclear structure and nuclear astrophysics. Following an overview of the most relevant achievements, we focus on the latest results on the $\beta$ decay of $^{60}$Ge and $^{62}$Ge. We also summarise our results on the mass excesses in comparison with systematics and a recent measurement. Finally, we present updated half-life trends for $T_z=$ -1/2, -1 and -2 neutron-deficient nuclides. 
}
\maketitle
\section{Introduction}
\label{intro}
Decay spectroscopy experiments with implanted radioactive ion beams (RIBs) are a powerful tool for the study of the structure of exotic nuclei. They provide rich spectroscopic information \mbox{\cite{Orrigo2014, Orrigo2016, OrrigoPRC2, Kucuk2017, Orrigo2018, Orrigo2021}:} firstly the half-life of the $\beta$-decaying nucleus, secondly the branching ratios for $\beta$-delayed $\gamma$ or particle emissions (where the particle can be a proton, an $\alpha$ particle, one or two neutrons, etc., depending on whether the unstable nucleus is neutron-deficient or neutron rich), thirdly information on the energy levels populated in the daughter nucleus together with their $\beta$-feedings and, last but not least, the absolute values of the Fermi \textit{B}(F) and Gamow-Teller \textit{B}(GT) transition strengths. In some cases one is also able to determine the mass of the daughter nucleus and estimate the mass of the parent. Furthermore, the structural information on exotic nuclei is relevant for nuclear astrophysics since many of these nuclei are involved in one of the processes of nucleosynthesis. Focusing on the neutron-deficient side of the nuclear chart around the $N=Z$ line, many heavy proton-rich elements are produced in the rapid proton-capture process (\textit{rp-process}) happening in explosive stellar environments, passing through neutron-deficient nuclei in the \textit{fp}-shell and above.

We have carried out a systematic study of neutron-deficient nuclei in $\beta$-decay spectroscopy experiments with implanted RIBs at GANIL and RIKEN obtaining remarkable results \cite{Orrigo2014, Orrigo2016, OrrigoPRC2, Kucuk2017, Orrigo2018, Orrigo2021}. Among the results from the GANIL experiments we emphasise the discovery of a new exotic decay mode in the \textit{fp}-shell in $^{56}$Zn ($\beta$-delayed $\gamma$-proton decay \cite{Orrigo2014}) and the first observation of the 2$^+$ isomer in $^{52}$Co \cite{OrrigoPRC2}. For some of the nuclides under study \cite{Orrigo2014, Orrigo2016, Orrigo2018} it was possible to enrich the $\beta$-decay data by comparison with complementary ($^3$He,\textit{t}) charge exchange (CE) reactions performed on the mirror stable target at RCNP Osaka \cite{Fujita2011, HFujita2013, Ganioglu2016}. CE reactions are the mirror strong interaction process of $\beta$ decay and provide information on relative \textit{B}(GT) values without energy restrictions \cite{Fujita2005, Fujita2011}, allowing us to investigate fundamental questions such as isospin symmetry in mirror nuclei.

The unprecedented statistics available at the Radioactive Isotope Beam Factory (RIBF) of the RIKEN Nishina Center allowed us to extend these studies to higher masses and more extreme nuclear conditions.
In Sect.~\ref{sec1} we focus on the most recent results on the $\beta$ decay of $^{60}$Ge and $^{62}$Ge obtained from the RIKEN experiment \cite{Orrigo2021}. Sect.~\ref{sec2} summarises our results on the mass excesses and compares them with the mass evaluation systematics and a recent measurement \cite{Paul2021}. In Sect.~\ref{sec3} we present half-life trends for $T_z=$ -1/2, -1 and -2 nuclei, where we include all the newly measured values. 

\section{$\beta$ decay of $^{60}$Ge and $^{62}$Ge}
\label{sec1}
\begin{figure}[h]
\centering
\includegraphics[width=0.7\columnwidth]{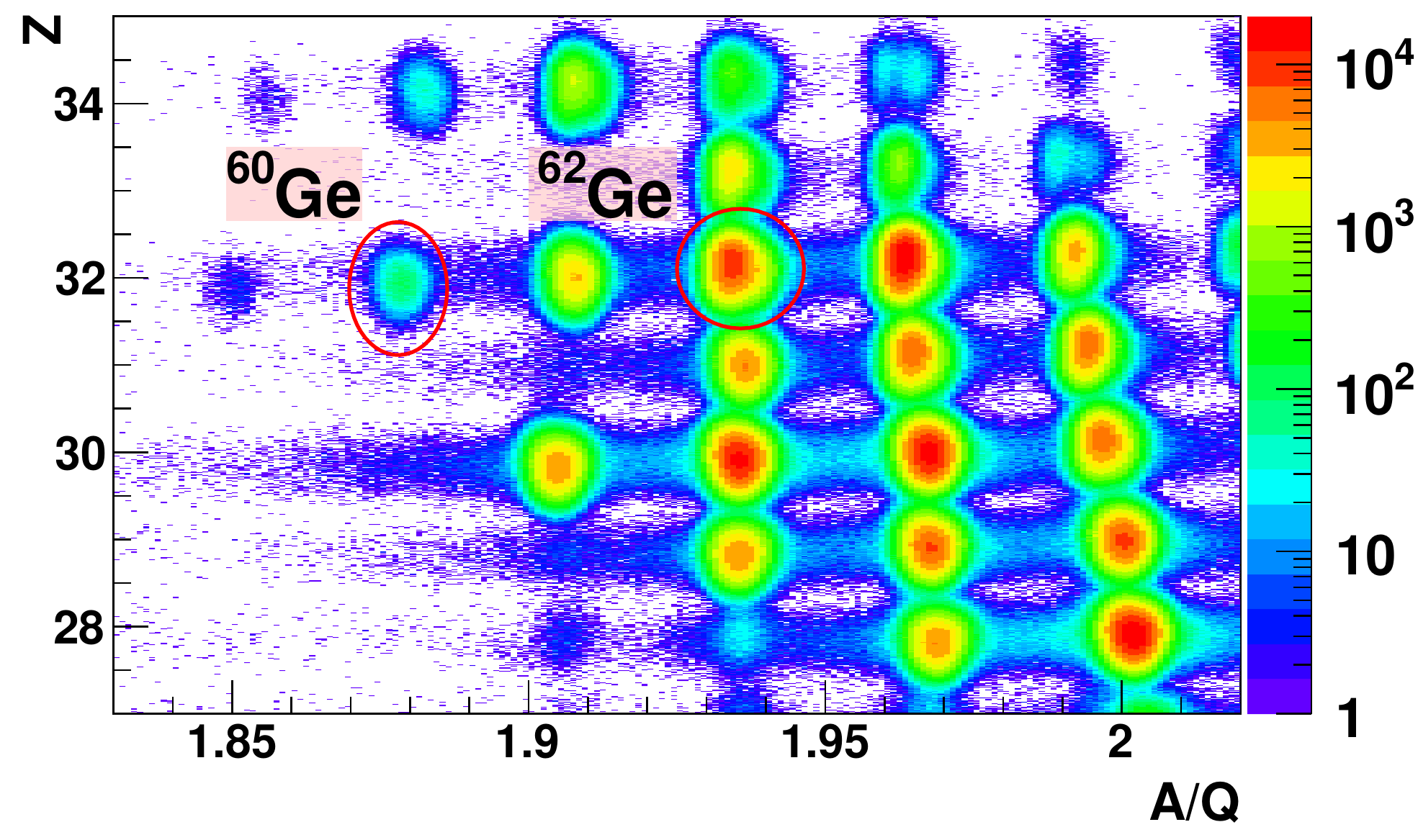}
\caption{Identification plot for the $^{60}$Ge and $^{62}$Ge implants.}
\vspace{-5 mm}
\label{fig1}
\end{figure}

$^{60}$Ge  and $^{62}$Ge were produced at RIBF by fragmentation of a $^{78}$Kr primary beam, accelerated to 345 MeV/nucleon with an intensity up to 250 pnA, on a Be target. The BigRIPS separator was used to select and identify the fragments with the $B\rho - \Delta E - ToF$ method \cite{Fukuda2013}. The isotopes were implanted in the WAS3ABi setup, using three double-sided Si strip detectors 60 mm $\times$ 40 mm $\times$ 1 mm. WAS3ABi was surrounded by the EURICA array (12 clusters containing 7 high-purity Ge crystals each) employed for $\gamma$ detection. The identification matrix is shown in figure~\ref{fig1}, where the positions of $^{60}$Ge and $^{62}$Ge are indicated. 1.5$\times$10$^4$ $^{60}$Ge and 2.1$\times$10$^6$ $^{62}$Ge ions were recorded \cite{Orrigo2021}.

$^{60}$Ge is a semi-magic, \mbox{$N$ = 28} isotone with \mbox{$T_z$ = -2} whose decay was almost unknown before the present experiment, as were the energy levels of its daughter nucleus $^{60}$Ga. The latter lies right at the proton drip-line, thus its structural properties are of relevance for the rp-process \cite{Paul2021}. Here a half-life value of $T_{1/2}$ = 25.0(3) ms has been measured for $^{60}$Ge as well as the first experimental information on both the $\beta$-delayed proton and $\gamma$ emissions \cite{Orrigo2021}. 7.6$\times$10$^5$ implants of the $^{60}$Ga ions were also observed and we obtained a half-life of 69.4(2) ms for $^{60}$Ga. The precision of both the $^{60}$Ge and $^{60}$Ga half-lives has been improved in comparison with values in the literature. In the $\beta$ decay of $^{60}$Ge we have also observed competition between the $\gamma$ de-excitation and the (isospin-forbidden) proton emission from the \mbox{$T$ = 2} isobaric analogue state (IAS) populated in the daughter nucleus. This exotic feature was already seen in all the lighter \mbox{$T_z=-2$} systems previously studied \cite{Orrigo2014, Orrigo2016, Orrigo2018}. 

Little was known about the $\beta$ decay of the \mbox{$T_z$ = -1}, $^{62}$Ge nucleus. Furthermore, $^{62}$Ge is of particular interest because in previous studies of \mbox{$T_z$ = -1} nuclei a suppression of isoscalar $\gamma$ transitions between \mbox{$J^{\pi}$ = 1$^+$}, $T$ = 0 states (Warburton and Weneser \textit{quasi-rule} \cite{Morpurgo58, Wilkinson69}) has been observed \cite{Molina2015}. A half-life value of \mbox{$T_{1/2}$ = 73.5(1)} ms has been extracted for $^{62}$Ge \cite{Orrigo2021}. This new value agrees with and improves the precision of our previous measurement (\mbox{$T_{1/2}$ = 76(6)} ms, obtained with 6.1$\times$10$^3$ implants of $^{62}$Ge \cite{Kucuk2017}). New information on the $\beta$-delayed $\gamma$ emission has been obtained, indicating the persistence of the \textit{quasi-rule} \cite{Morpurgo58, Wilkinson69} in $^{62}$Ge. Finally, we do not find evidence of enhanced low-lying Gamow-Teller strength in $^{62}$Ga due to isoscalar proton-neutron pairing, confirming the findings of a previous measurement~\cite{Grodner2014}.

\begin{figure}[t!]
\centering
\includegraphics[width=0.6\columnwidth]{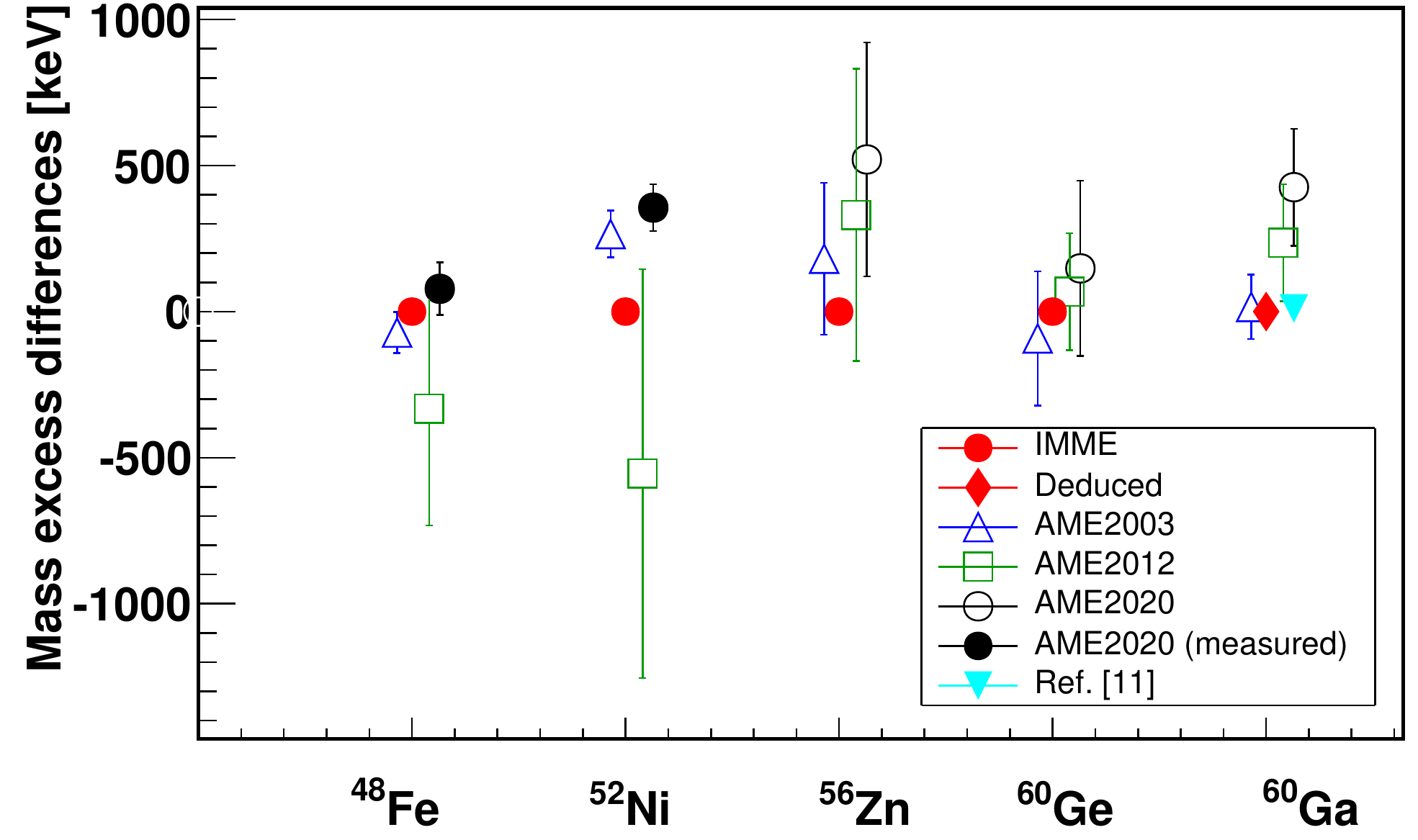}
\caption{Mass excesses of $^{48}$Fe, $^{52}$Ni, $^{56}$Zn, $^{60}$Ge and $^{60}$Ga. The difference is shown between the values we obtain (red filled circles and diamond) \cite{Orrigo2016, Orrigo2021} and values from the 2003, 2012 and 2020 AMEs \cite{Audi2003, Audi2012, Wang2021}. Open symbols represent values from systematics, while filled symbols are experimental values. The cyan filled triangle is the measurement from Ref. \cite{Paul2021}. The data points belonging to each nucleus are slightly displaced to show the error bars better.}
\vspace{-5 mm}
\label{fig2}
\end{figure}

\section{Mass excess of neutron-deficient nuclei}
\label{sec2}
Knowing the mass excesses of the neutron-deficient nucleus undergoing $\beta$ decay and its daughter nucleus is key to determining important quantities such as the $Q_{\beta}$ value of the decay which, in turn, enters in the determination of \textit{B}(F) and \textit{B}(GT). Moreover, the information on the mass excesses of the $\beta$- and $\beta$-proton daughter nuclei allows the calculation of the proton separation energy \textit{S$_p$}. Since there is a lack of mass measurements in this mass region, one has to rely often on the atomic mass evaluation (AME) systematics. However, if the mass excesses of at least three members of an isospin multiplet are known, another option is to determine the mass excess of the remaining member of the multiplet from the Isobaric Multiplet Mass Equation (IMME) \cite{IMMEpaper, IMME, MacCormick2014}:

\begin{equation}
  \Delta m(\alpha,T,T_z) = a + b~T_z + c~T_z^2\,.
  \label{Eq1}
\end{equation}
\\
where $T_z$ is the third component of the isospin $T$ and $\alpha$ stands for all the other quantum numbers. We used our $\beta$-decay data to determine the ground-state (g.s.) mass excesses of the \mbox{$T_z$ = -2} nuclei $^{48}$Fe, $^{52}$Ni and $^{56}$Zn in Ref. \cite{Orrigo2016} and $^{60}$Ge in Ref. \cite{Orrigo2021} from the IMME, knowing four members of each quintuplet. We have also deduced the g.s. mass excess of $^{60}$Ga and its \textit{S$_p$} \cite{Orrigo2021}. In figure~\ref{fig2} our deduced mass excesses (red filled circles and diamond) are compared to the values obtained from the 2003 \cite{Audi2003}, 2012 \cite{Audi2012} and 2020 \cite{Wang2021} AME systematics. In the figure open (filled) symbols represent the AME systematics (measured or deduced) values. 

Since our first study of $^{56}$Zn \cite{Orrigo2014} we have observed that for proton-rich nuclei in this region of the mass chart the AMEs subsequent to that in 2003 give a worse agreement for the mass excess in comparison with our IMME values. In figure~\ref{fig2} one can see that the values from the 2003 AME lie closer to our estimates than the values from the 2012 AME and a very similar behaviour happens for the 2016 AME \cite{Wang2017} (not shown in the figure). Other authors then reported similar issues \cite{DelSanto2014, Paul2021}. The 2020 AME also has a similar behaviour for the unmeasured nuclei, but includes new measured values for $^{48}$Fe and $^{52}$Ni (black filled circles). The AME systematic evaluation does not include isobaric multiplets. The reasoning is that the IAS might be mixed and consequently its energy can deviate from the isobaric multiplet formula. The $^{56}$Zn IAS is indeed mixed  \cite{Orrigo2014}. We think that, with caution, this knowledge can help to calculate extrapolated values. The future AME would benefit from new mass measurements in this region. 

Figure~\ref{fig2} also shows a recent measurement of the mass excess in $^{60}$Ga (cyan filled triangle) \cite{Paul2021}, in excellent agreement with our indirect determination based on the $\beta$-decay data \cite{Orrigo2021}. Paul \textit{et al.} have also determined \textit{S$_p$}= 78(30) keV in $^{60}$Ga, in agreement with our value of 90(15) keV \cite{Orrigo2021} (the 2020 AME systematic value is -340(200) keV \cite{Wang2021}) and combined the two experimental values to give \textit{S$_p$}($^{60}$Ga) = 88(18) keV, establishing the proton-bound nature of $^{60}$Ga. Then, since $^{59}$Ga was not observed in fragmentation reactions at NSCL, there is strong evidence that $^{60}$Ga is the last proton-bound gallium isotope \cite{Paul2021}. Our recent experiment at RIKEN confirms that $^{59}$Ga is not observed, providing further evidence that $^{60}$Ga marks the location of the proton drip line.

\section{Half-life trends}
\label{sec3}
\begin{figure}[ht]
\centering
\sidecaption
\includegraphics[width=0.6\columnwidth]{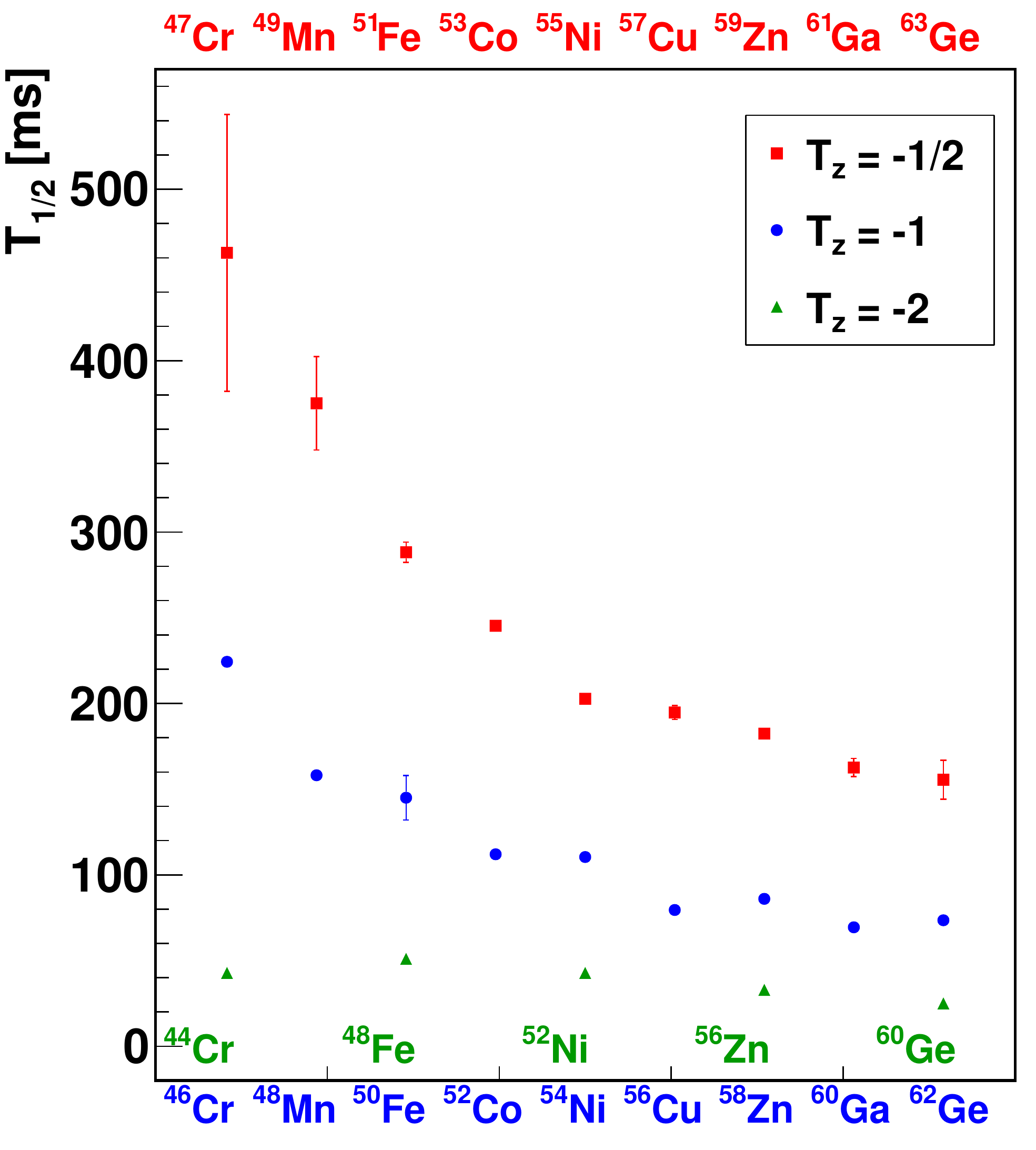}
\caption{$T_{1/2}$ values measured for the $T_{z}$ = -1/2 (red squares), -1 (blue dots) and -2 (green triangles) nuclei as~a function of atomic number.}
\label{fig3}
\end{figure}

In figure 7 from Ref. \cite{Kucuk2017} we have analysed the trend observed in the half-lives as a function of atomic number for the $T_{z}$ = -1/2 and -1 nuclei. Here, in figure~\ref{fig3}, we present a new version of that figure, where we have updated the half-life values and included the trend for \mbox{$T_{z}$ = -2} nuclei. In particular, we have included/updated the half-lives of the following nuclides: $^{56}$Zn \cite{Orrigo2014}; $^{48}$Fe and $^{52}$Ni \cite{Orrigo2016}; $^{52}$Co \cite{OrrigoPRC2}; $^{60}$Ge, $^{60}$Ga, $^{62}$Ge and $^{59}$Zn \cite{Orrigo2021} and $^{44}$Cr \cite{Dossat2007}.

As discussed in Ref. \cite{Kucuk2017}, the systematic decrease of the $T_{1/2}$ values with the mass, observed in the $T_{z}$ = -1/2 nuclei, reflects the increase in the $Q_\beta$ value. A similar decreasing pattern is also found in the half-lives of the $T_{z}$ = -1 nuclei and, on top of this behaviour, a typical odd-odd and even-even effect is observed. This is due to the fact that in the odd-odd nuclei there exist other excited states below the IAS that receive a significant amount of $\beta$ feeding, which makes their half-lives slightly shorter in comparison with their even-even neighbours \cite{Kucuk2017}. In the $T_{z}$ = -2 nuclei, which are all even-even, a similar decreasing trend is observed, with the half-life of the most exotic $^{60}$Ge being only 25.0(3) ms. 

\section{Conclusions}
\label{conc}
A panoramic view of the most relevant achievements from our $\beta$-decay spectroscopy experiments, with a focus on the most recent results, has been presented. The half-life trends as a function of the mass number and a comparison of our deduced mass excesses with different AME systematics have also been discussed. Altogether, it is a rich example of the valuable spectroscopic information that can be obtained from this kind of experiment.
\\

\begin{acknowledgement}
\textbf{Acknowledgments}. This work was supported by the Generalitat Valenciana Grant No. PROMETEO/2019/007, Spanish Grants No.~PID2019-104714GB-C21, FPA2017-83946-C2-1-P and FPA2014-52823-C2-1-P (MCIN,MINECO/AEI/FEDER), Centro de Excelencia Severo Ochoa del IFIC SEV-2014-0398; \textit{Junta para la Ampliaci{\'o}n de Estudios} Programme (CSIC JAE-Doc) co-financed by FSE.
\end{acknowledgement}
%
%
%

\end{document}